\documentclass[11pt]{article}

\usepackage{amssymb}
\usepackage{graphicx}
\title{Relative equilibria of four identical satellites}

\date{}

\begin{document}

\maketitle

\centerline {Alain Albouy, albouy@imcce.fr}
\centerline {CNRS-UMR 8028, Observatoire de Paris}
\centerline {77, avenue Denfert-Rochereau}
\centerline {75014 Paris, France}
\bigskip
\centerline{Yanning Fu, fyn@pmo.ac.cn}
\centerline{Purple Mountain Observatory}
\centerline{2 West Beijing Road}
\centerline{Nanjing 210008, P.\ R.\ China}
\bigskip

\begin{abstract}
We consider the Newtonian 5-body problem in the plane, where 4 bodies have the same mass $m$, which is small compared to the mass $M$ of the remaining body. We consider the (normalized) relative equilibria in this system, and follow them to the limit when $m/M\to 0$. In some cases two small bodies will coalesce at the limit. We call the other equilibria the relative equilibria of four separate identical satellites. We prove rigorously that there are only three such equilibria, all already known after the numerical researches in [SaY]. Our main contribution is to prove that any equilibrium configuration possesses a symmetry, a statement indicated in [CLO2] as the missing key to proving that there is no other equilibrium.
\end{abstract}

\bigskip

\centerline{\bf 1. Introduction}

The recent  work by Renner and Sicardy ([SiR], [ReS], [Ren]) attracted
again the attention of astronomers on relative equilibria in the
$N$-body problem, specially those with a big mass and several small masses. In such a relative equilibrium the small bodies all describe almost the same circular
orbit around the central body. Some of these relative equilibria are
stable, and thus, may approximate the motion of small bodies in the
solar system. Actually, according to a conjecture by Moeckel [Moe],
the existence of a dominant mass could be a necessary condition for
the stability of a relative equilibrium in the $N$-body problem.

Let us recall briefly the previous work. Lagrange [Lag] discovered
the famous relative equilibrium with three bodies forming an
equilateral triangle. Gascheau [Gas] gave the condition for
stability, which implies that the mass of one of the bodies should
be at least 25 times the total mass of the other bodies. The Trojan
asteroids were discovered in 1906, and, more recently, many other
configurations in equilateral triangle were found (see e.g.\ [ReS], [RoS]).
While studying Saturn's rings, Maxwell [Max] proposed a
configuration with many small bodies with equal masses orbiting on
the same circle and forming a regular polygon, and proved its
stability. He also examined the more realistic case where the small
bodies have different masses and the polygon is not regular, and
claimed again the stability. This sort of thin discrete ring has not
been observed yet. Maxwell's stability claims were reconsidered, and
confirmed for the regular polygon with 7 vertices or more ([Moe],
[Rob]).

Several models were proposed (see [SaH], [NaP]) to explain the four
or five co-orbital arcs of ring around Neptune. Renner and Sicardy
(see also the next to last paragraph in [NaP]) suggest that three or
four little unknown satellites would orbit in a relative equilibrium
configuration, and control the ends of the arcs. The equilibrium of
the system composed of the arcs and the satellites would be stable
enough and compatible with the close inner satellite Galatea ([Ren]).

When speaking of stability above, we meant linear stability of the
relative equilibrium. A relative equilibrium is a motion where the
distances between bodies remain the same. The relative motion of
any body around any other is circular. All the bodies orbit
in the same plane, but the linear stability includes small
displacements of the positions and the velocities out of this plane.

When a configuration may have a relative equilibrium motion, it may
also have a so-called homographic motion. Here the configuration
changes size, keeping the same shape, each body being in the same
plane and describing a Keplerian orbit around each other. So, one
should discuss not only the stability of the circular motion, i.e.\
the relative equilibrium, but also the stability of the elliptic motion
(see [MSS], [Rob2], [MeS], [HuS]).

It is good to give a name to the configurations allowed in relative
equilibria and homographic motions. They are called the central
configurations of the $N$-body problem. The limiting configurations where the big mass is fixed, while the $n=N-1$ small masses tend to zero together, their mutual ratios being fixed, are called the central configurations of the $1+n$-body problem.

Our work as well as several previous studies is dedicated to the
{\sl non-coales\-cent planar central configurations of the $1+n$-body
problem with small equal masses}. All the words are required. {\sl
Non-coalescent} excludes the case where two small bodies coincide at
the limit (see [Xia] and [Moe2]). {\sl Planar} should be indicated,
because central configurations may also be non-planar. A special
type of homographic motion, the homothetic motion, does not
require that the configuration is in a plane. In a homothetic motion
the configuration is central and does not rotate.

We will avoid such a heavy terminology considering it as equivalent
as {\sl relative equilibria of $n$ separate identical satellites}.
A relative equilibrium is not only a configuration. Velocities
should be attributed to the bodies in order to obtain a uniform
rotation. But the choice of the velocities being mainly
unique, the distinction between relative equilibrium and central
configuration does not matter here.

Hall [Hal] gave a precise proposition proving in particular that in
a relative equilibrium of $n$ satellites, the big body is at the
center of a circle which passes through the satellites. In other
words, the satellites are co-orbital. A couple of years after Hall's
unpublished work, Salo and Yoder [SaY] studied independently the
relative equilibria of $n$ separate identical satellites, and
discovered that they form strange sequences. When $n=2, 3,\dots,10,\dots$
there are respectively $2,3,3,3,3,5,3,1,1,\dots$ distinct such
equilibria. Thus, very probably, from $n=9$, there is only one
equilibrium, which is the regular polygon. All the equilibria in
[SaY] list have exactly one axis of symmetry (passing through the
central body), except the regular polygons, which have more
symmetry. The number of linearly stable relative equilibria in [SaY]
list is, again for $n=2,3,\dots$, respectively
$1,1,1,1,1,2,2,1,1,\dots$

The numerical experiments suggest several mathematical statements.
Very few of them are proved. Hall proved rigorously that there
are 3 relative equilibria of $3$ separate identical satellites. He then claimed: ``For $n\geq 4$ the equations become considerably more difficult to handle. We will give some (little) numerical description for $n\geq 3$ in section 7."

Improving another result by Hall, Casasayas, Llibre and Nunes [CLN]
proved that the regular polygon is the unique equilibrium if $n>e^{73}$.

Cors, Llibre and Olle proved that there are only three symmetric equilibria of 4 separate identical satellites. We prove here that any such equilibrium must be
symmetric, thus completing the proof of the following

{\bf Theorem.} There are exactly 3 relative equilibria of 4
separate identical satellites. In the corresponding configurations ${\cal E}_1$, ${\cal E}_2$, ${\cal E}_3$,
the small bodies are placed on a circle with center the big body, in
such a way that the four angular distances between consecutive
bodies are respectively $(\pi/3,2\pi/3,2\pi/3,\pi/3)$,
 $(\theta_1,\theta_2,\theta_1,\theta_4)$, $(\pi/2,\pi/2,\pi/2,\pi/2)$, where $\theta_1$, $\theta_2$, $\theta_4$ satisfy $2\theta_1+\theta_2+\theta_4=2\pi$, $0<\theta_2<\theta_1<\pi<\theta_4$.

\begin{figure}[t]\label{ccfg}
\vspace{1cm}\centerline{\includegraphics [width=100mm] {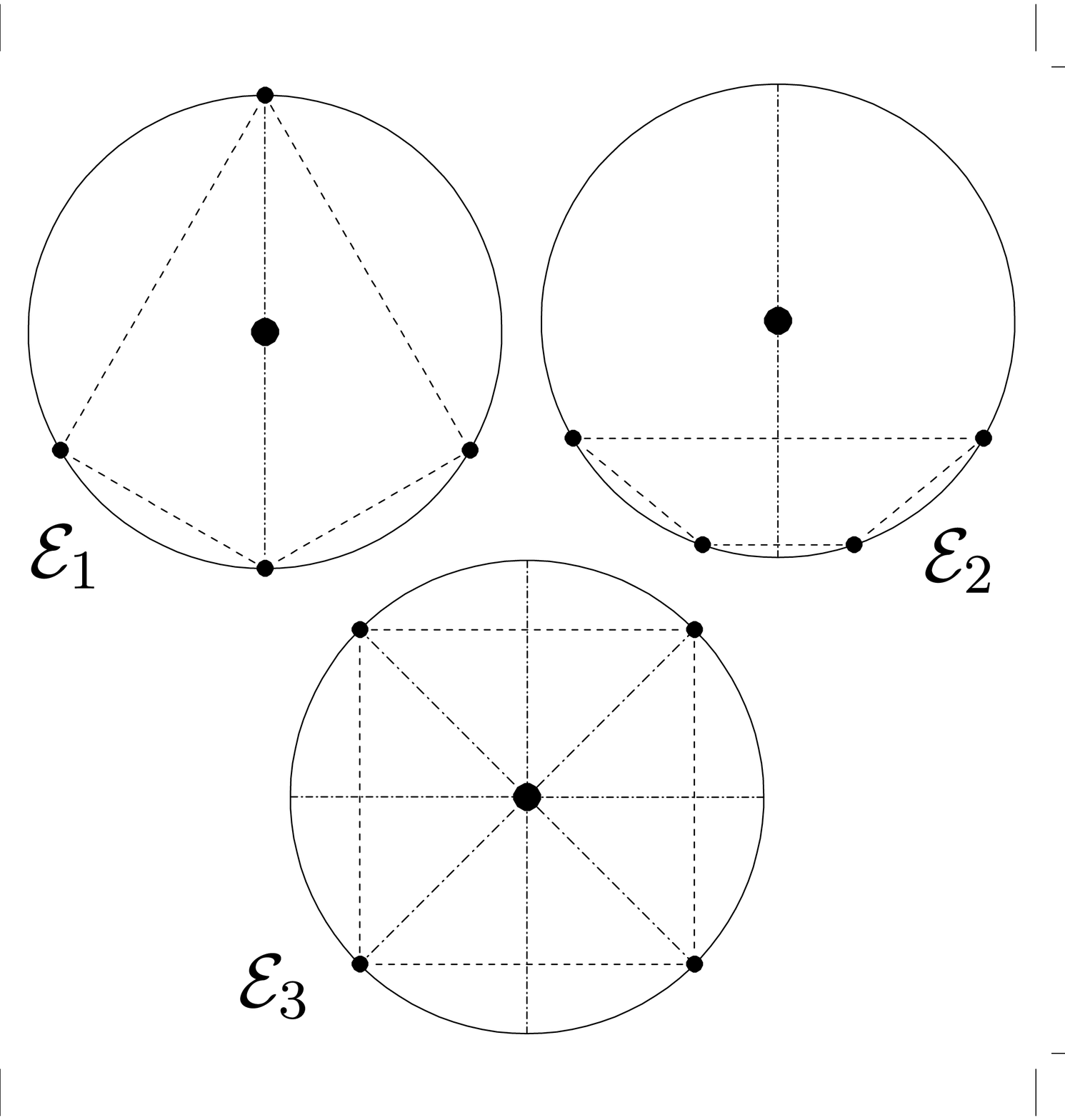}}
\vspace{0cm} \caption{The three relative equilibria of 4 separate
identical satellites}
\end{figure}

Let us discuss briefly the most natural extensions of our Theorem, first removing the condition ``separate". According to Mart\'\i nez and Sim\'o [MS], there are exactly 8 other relative equilibria of four identical satellites, counting configurations of indistinguishable satellites up to isometry. Three of them are 2-dimensional and asymmetric, two are 2-dimensional and have one axis of symmetry, and three are 1-dimensional (Moulton configurations).

Renner and Sicardy [ReS] removed the condition ``identical" and obtained many results, including mathematical results about the inverse problem: given a configuration of the satellites, find the small masses making it a relative equilibrium.

Finally, the non-coalescent 3-dimensional central configurations of the $1+4$-body problem with equal small masses were also studied. Albouy and Llibre [AlL] proved their symmetry, but left open the possibility of (infinitely) many configurations with exactly one plane of symmetry. Numerical studies show there is only one such central configuration. The configurations discussed in [AlL] being non-planar, they are not allowed in a motion of relative equilibrium. However, together with
the planar central configurations classified by the Theorem above,
they would complete the classification of homothetic motions with a
big body and 4 bodies with equal small enough masses (see [Rob] and
[Alm] for examples of explicit versions of the ``small enough" condition).

\bigskip
\centerline{\bf 2. Study of a function}

The equations characterizing the relative equilibria of four
identical satellites are well-known (see e.g.\ [CLN], [CLO], [Hal],
[Moe], [SaY]). In such an equilibrium, the four bodies with small
mass lie on a circle centered at the body with big mass. So one considers only the normalized configurations with the same property when formulating the equations. The variables are the four positive angular distances between two consecutive small bodies. These four angles $\theta_1$, $\theta_2$, $\theta_3$, $\theta_4$ satisfy $\theta_1+\theta_2+\theta_3+\theta_4=2\pi$.

We define the function
\begin{equation}\label{fun}
f(\theta)=\sin\theta\bigl(1-(2-2\cos\theta)^{-3/2}\bigr),
\end{equation}
and set $$f_1=f(\theta_1),\dots,f_4=f(\theta_4),$$
$$f_{12}=f(\theta_1+\theta_2),f_{23}=f(\theta_2+\theta_3),f_{34}=f(\theta_3+\theta_4),
f_{14}=f(\theta_1+\theta_4).$$ Obviously $f_{14}=-f_{23}$,
$f_{12}=-f_{34}$. The equations are:
\begin{equation}\label{cc}
f_{34}=f_2-f_3=f_1-f_4,\qquad f_{23}=f_1-f_2=f_4-f_3.
\end{equation}

In this section we state some properties of $f$ which, together with some simple inequalities involving $f(\pi/6)$, $f(\pi/2)$ and $f'(\pi/2)$, are necessary and sufficient for us to
prove the symmetry of the central configurations. As
$f^{(n)}(2\pi-\theta)=(-1)^{n+1}f^{(n)}(\theta)$, it is enough to focus on $\theta \in ]0,\pi]$.

\begin{figure}\label{f}
\vspace{1cm}\centerline{\includegraphics [width=100mm] {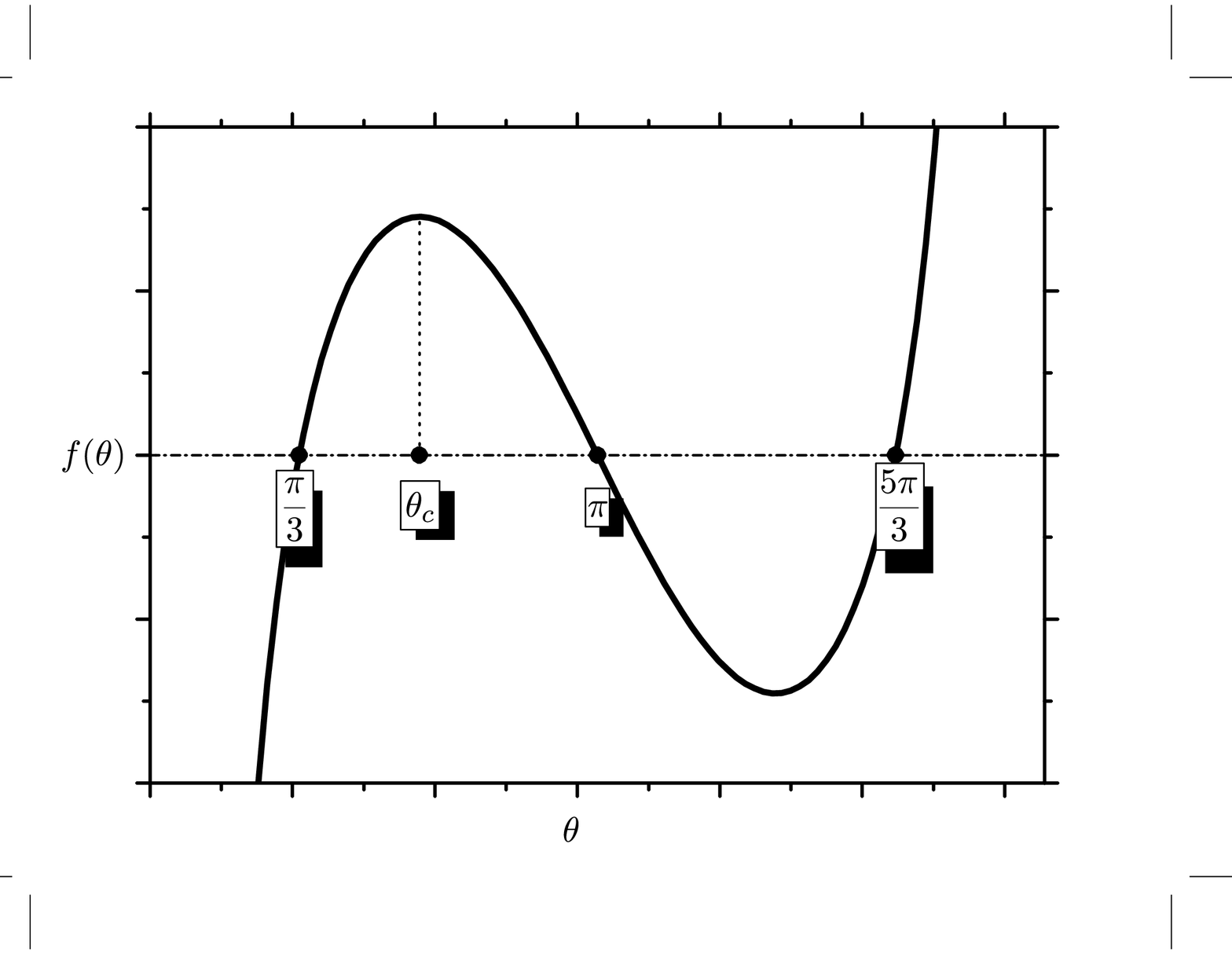}}
\vspace{0cm} \caption{The function $f$}
\end{figure}

{\sl Property 1.} We have $f(\pi/3)=f(\pi)=0$. If
$\theta\in]0,\pi/3[$ then $f(\theta)<0$. If $\theta\in]\pi/3,\pi[$
then $f(\theta)>0$.

{\sl Property 2.} In $]0,\pi[$ there is a unique $\theta_c$ such that $f'(\theta_c)=0$, satisfying $\theta_c>\pi/3$. If $\theta\in]0,\theta_c[$, $f'(\theta)>0$, if $\theta\in]\theta_c,\pi[$, $f'(\theta)<0$.

{\sl Property 3.} We have $f''(\pi)=0$, and  $f''(\theta)<0$ in $]0,\pi[$.

{\sl Property 4.} [Hal] We have $f'''(\theta)>0$ in $]0,\pi]$.

{\sl Property 5.} Consider $g:]f(0^+),f(\theta_c)[\to]0,\theta_c[$
such that $g(f(\theta))=\theta$, which is well-defined according to Property 2.
We have $g'''>0$.

{\sl Proof.} The identity $2\sin^2(\theta/2)=1-\cos\theta$ suggests the introduction of some intermediate notation in the expression of $f$. We write
\begin{equation}\label{fun2}
f(\theta)=\sin\theta\bigl(1-(2s)^{-3}\bigr),\quad\hbox{with}\quad s=\sin{\theta\over 2},\quad\theta\in]0,2\pi[.
\end{equation}
We may notice that if $\theta\in]2\pi,4\pi[$ formulas (\ref{fun}) and (\ref{fun2}) disagree, because $s<0$.
But we are only interested in the interval $]0,2\pi[$. We find
$$f'(\theta)={1\over 4s^3}-{1\over 8s}+1-2s^2,\quad f''(\theta)=\sin\theta(-{3\over 16s^5}+{1\over 32s^3}-1),$$
$$f'''(\theta)={3\over 4s^5}-{5\over 8s^3}+{1\over 32s}-1+2s^2.$$
The rational substitution
\begin{equation}\label{rat}
s=\frac{\alpha_0+\alpha_1 y}{\beta_0+\beta_1 y}
\end{equation}
mapping $y\in ]0,\infty[ ~ \mapsto s \in
]\alpha_0/\beta_0,\alpha_1/\beta_1[$ is often efficient in proving
rigorously that a rational function in $s$ is sign definite in an
interval. Taking $\alpha_0=0,\alpha_1=1,\beta_0=1,\beta_1=1$, this
substitution reads $s=y/(1+y)$ and gives
$32(1+y)^2y^5f'''=37y^7+7y^6+275y^5+641y^4+740y^3+484y^2+168y+24$,
so $f'''$ is positive when $y\in]0,+\infty[$, i.e.\ $s\in]0,1[$,
i.e.\  $\theta\in ]0,2\pi[$. This proves Property 4. We deduce
immediately Properties 1 to 3, after checking
$f''(\pi)=f(\pi)=f(\pi/3)=0$.

We consider now Property 5. We differentiate twice the relation
$g(f(\theta))=\theta$, substitute $g'(f(\theta))$, differentiate
again and find, after substituting $g''(f(\theta))$,
$f'^5(\theta)g'''(f(\theta))=-f'(\theta) f'''(\theta) + 3
f''^2(\theta)$. We compute this expression, substitute $s=y/(1+y)$
and get $256y^8(1+y)^4f'^5(\theta)g'''(f(\theta))=
259y^{12}+7412y^{11}+16934y^{10}+32960y^9+42564y^8+39236y^7+35306y^6+32904y^5+24389y^4+12208y^3+3880y^2+720y+60$.
This shows that $g'''(f(\theta))$ is positive if
$\theta\in]0,\theta_c[$. $\Box$

{\bf Lemma 1.} If a function $\phi$ satisfies $\phi'''>0$ on the interval
$]a,b[$, then for any $t_1$, $t_2$, $t_3$, $t_4$, $a \leq t_1< t_2<
t_3< t_4\leq b$, we have $$\left|\matrix{1&1&1&1\cr
t_1&t_2&t_3&t_4\cr t_1^2&t_2^2&t_3^2&t_4^2\cr
\phi(t_1)&\phi(t_2)&\phi(t_3)&\phi(t_4)}\right|>0.$$

{\sl Remark.} A more general statement is stated and proved as Problem 5-99 in [PoS].

{\bf Lemma 2.} Consider two horizontal chords drawn on the graph of
$f$, whose projections on the horizontal axis are the two segments
$[t_1^L,t_1^R]$, $[t_2^L,t_2^R]$, with
$t_1^L<t_2^L<\theta_c<t_2^R<t_1^R$. Then $t_2^L+t_2^R<t_1^L+t_1^R$,
that is, the segment from the middle point of one chord to the
middle point of the other chord has a negative slope.

{\sl Proof.} By Property 4, $f$ satisfies the hypothesis of Lemma 1,
so, setting $f_1=f(t_1^L)=f(t_1^R)$, $f_2=f(t_2^L)=f(t_2^R)$,
we get $$\left|\matrix{1&1&1&1\cr t^L_1&t^L_2&t_2^R&t_1^R\cr {t^L_1}^2&{t^L_2}^2&{t_2^R}^2&{t_1^R}^2\cr f_1&f_2&f_2&f_1}\right|>0.
$$ The
determinant factorizes as
$(f_2-f_1)(t_1^R-t_1^L)(t_2^R-t_2^L)(t_1^L+t_1^R-t_2^L-t_2^R)$.
$\Box$

{\bf Corollary.} For any horizontal chord as in Lemma 2, projecting on the segment $[t^L,t^R]$,
$t^L<\theta_c$, we have $2\theta_c<t^L+t^R$.

{\sl Proof.} This is Lemma 2 when the highest chord tends to the horizontal tangent at $(\theta_c,f(\theta_c))$. $\Box$

{\bf Lemma 3.} Consider two chords drawn on the first increasing
branch of the graph of $f$, the arc delimited by the ``inner" chord
being included in the arc delimited by the ``outer" one. If the
segment from the middle of one chord to the middle of the other
chord is horizontal, then the slope of the outer chord is less than
the slope of the inner chord.

{\sl Proof.} By Property 5 we can apply Lemma 1 to the reciprocal
function $g$ of $f$. We take some notation. The outer chord projects
horizontally on the segment $[t_1,t_4]$, the inner one on the
segment $[t_2,t_3]$, with $t_1<t_2<t_3<t_4<\theta_c$. We set
$f_i=f(t_i)$, $i=1,\dots, 4$ and get $$\left|\matrix{1&1&1&1\cr
f_1&f_2&f_3&f_4\cr f_1^2&f_2^2&f_3^2&f_4^2\cr
t_1&t_2&t_3&t_4}\right|>0$$ By the hypothesis $f_1+f_4=f_2+f_3$ the
determinant factorizes as
$(f_4-f_3)(f_3-f_1)\bigl((t_4-t_1)(f_3-f_2)-(t_3-t_2)(f_4-f_1)\bigr)$,
so we have $(t_4-t_1)(f_3-f_2)-(t_3-t_2)(f_4-f_1)>0$. $\Box$

{\sl Remark 1.}  A result similar to Lemma 3 was used in [AlV] to prove the symmetry of central configurations in the $4+1$-body problem, the four equal big masses forming a square.

{\sl Remark 2.} We could consider the general homogeneous law of
force instead of Newton's force. Newton's force would then be the
particular case ${p}=-3$ in the expressions
\begin{equation}\label{fp}
f(\theta)=\sin\theta(1-(2s)^{{p}}),
\end{equation}
\begin{equation}\label{fpp}
f'(\theta)=1-2s^2-(2s)^{p}(1+{p})+(2+{p})s^2(2s)^{p},
\end{equation}
\begin{equation}\label{fppp}
f''(\theta)=\sin\theta(-{p}(1+{p})(2s)^{{p}-2}+(2+{p})^22^{{p}-2}s^{p}-1).
\end{equation}
The problem of point vortices on a plane [ONe] would correspond to
${p}=-2$. It may be asked if the results presented here are still
true for other values of ${p}$. It seems that there are always 3
relative equilibria for any $p<0$, but our arguments does not
cover all this interval. Only Property 1 is true for any negative
$p$. Property 2 requires ${p}\in]-\infty,-1[$. According to
expression (\ref{fppp}), Property 3 is true on the interval
$]\infty,\rho]$, where $\rho=-1.0022967...$ is the root of the
equation\footnote{Here we use an easy generalization of the
classical discriminants $b^2-4ac$ and $4p^3-27q^2$. The expression
$aX^\alpha+bX^\beta+cX^\gamma$ has a double root if
$$\left(\frac{a}{\gamma-\beta}\right)^{\gamma-\beta}\left(\frac{b}{\alpha-\gamma}\right)^{\alpha-\gamma}
\left(\frac{c}{\beta-\alpha}\right)^{\beta-\alpha}=1.$$} in $p$
$$4(-p-1)^p\bigl((2+p)^2\bigr)^{2-p}=(8-4p)^{2-p}.$$ We exclude the
second case in Proposition 1 using Properties 1 to 3 and Property 5.
Numerical studies indicate that Property 5 is always true in the
previous interval $]\infty,\rho]$. We exclude the first case using
Properties 1 to 4. Property 4 imposes further restrictions.
According to numerical studies, we should restrict to
$p\in]-37.61045,-1.26766[$. To finish the classification of the
relative equilibria, we will need in particular some polynomial
calculus which can be carried out only for simple integer or
rational values of $p$.

\bigskip
\centerline{\bf 3. Symmetry result}

{\bf Proposition 1.} Consider any solution of (\ref{cc}), given by the four positive angles $\theta_i$ between the adjacent bodies, which satisfy $\theta_1+\theta_2+\theta_3+\theta_4=2\pi$. At least one of the following situations occurs: a diagonal of the quadrilateral is a diameter of the circle (i.e.\ $\theta_1+\theta_2=\pi$ or $\theta_2+\theta_3=\pi$), or  two angles among $\theta_1$, $\theta_2$, $\theta_3$ and $\theta_4$ are equal.

{\sl Proof.} We assume the conclusion is not satisfied and will
derive a contradiction. We can take without loss of generality the
labeling convention such that $\theta_4<\theta_2<\theta_1$ and
$\theta_3<\theta_1$.  This implies
$\theta_3+\theta_4<\theta_1+\theta_2$, which is
$\pi<\theta_1+\theta_2$. The remaining assumption is
$\theta_2+\theta_3\neq \pi$, that is, we are either in the {\sl
first case} $\theta_1+\theta_4<\pi$ or in the {\sl second case}
$\theta_2+\theta_3<\pi$.

{\sl First case.} Here $\theta_1<\theta_1+\theta_4<\pi$.
As $\theta_1$ is the greatest angle, $\pi/2<\theta_1$, thus $f_1>0$ and $f_{14}>0$. From the
last inequality and the equations (\ref{cc}), we get $f_3>f_4$
and $f_2>f_1$, the latter with $\theta_2<\theta_1$ implying
$\theta_c<\theta_1$.

By the relation  $f_{34}-f_{14}=f_1-f_3$, the expression
$A(\theta)=f(\theta_1+\theta)-f(\theta_3+\theta)$ is such that
$A(\theta_4)=-A(0)$. We have, for $\theta\in [0,\theta_4]$,
$A'(\theta)=f'(\theta_1+\theta)-f'(\theta_3+\theta)<0$,  by Property 3. So $A(0)>0$, that is $f_1>f_3$.
Together with what we obtained, this gives
$f_2>f_1>f_3>f_4$.

From $\theta_1>\theta_3$ and $f_1>f_3$, we know that $\theta_3<\theta_c$.
The order $\theta_2<\theta_3$ is then impossible, as it would imply
$f_2<f_3$. Also our case hypothesis is
$\theta_1+\theta_4<\theta_2+\theta_3$, or
$\theta_1-\theta_2<\theta_3-\theta_4$. Putting these arguments
together, we get the order $\theta_4<\theta_3<\theta_2<\theta_1$.

We know that $f_2>f_1>0$ and $\theta_c<\theta_1$. Let us call $\theta_1^L<\theta_c$  the unique angle satisfying this inequality such that $f(\theta_1^L)=f_1$. We call $\theta_2^L$ and
$\theta_2^R$, $\theta_2^L<\theta_c<\theta_2^R$, the two angles such
that $f(\theta_2^L)=f(\theta_2^R)=f(\theta_2)=f_2$. So we have
either $\theta_2=\theta_2^L$ or $\theta_2=\theta_2^R$. We get
\begin{equation}\label{c1}
0<\frac{f_2-f_1}{\theta_2^L-\theta_1^L}<\frac{f_3-f_4}{\theta_3-\theta_4}<\frac{f_2-f_1}{\theta_1-\theta_2}\leq\frac{f_2-f_1}{\theta_1-\theta_2^R}
\end{equation}
All the numerators take the same value
$f_2-f_1=f_3-f_4>0$. The first inequality between slopes is due to
$f''<0$. The second is our case hypothesis. The last
is simply $\theta_2\leq\theta_2^R<\theta_1$. Now, the inequality resulting from (\ref{c1}), namely
$\theta_1-\theta_2^R<\theta_2^L-\theta_1^L$, or $\theta_1^L+\theta_1<\theta_2^L+\theta_2^R$, contradicts Lemma 2.

{\sl Second case.} Here $\theta_2+\theta_3<\pi$. From the
combination $f_{34}-f_{23}=f_2-f_4$, exactly as we did in the
previous case, we deduce $f_2>f_4$ and $f_{34}>f_{23}$. The latter combined with $\theta_3+\theta_4<\theta_2+\theta_3$ implies
$\theta_c<\theta_{2}+\theta_3$.
As a particular result, we have $f_{23}>0$, and so, $f_1>f_2$ and
$f_4>f_3$. Putting together, we get $f_1>f_2>f_4>f_3$. This fits with the conventions
$\theta_3<\theta_1$ and $\theta_4<\theta_2<\theta_1$ only if $\theta_3<\theta_c$ and $\theta_4<\theta_2<\theta_c$. It is then obvious
from $f_3<f_4$ that $\theta_3<\theta_4$. This gives the order
$\theta_3<\theta_4<\theta_2<\theta_1$. Let $\theta_1^L<\theta_c$ be the unique angle satisfying this inequality such that $f(\theta_1^L)=f_1$. We have $\theta_3<\theta_4<\theta_2<\theta_1^L<\theta_c$.
According to (\ref{cc}), Lemma 3 may be applied, giving
$(\theta_1^L-\theta_3)(f_2-f_4)-(\theta_2-\theta_4)(f_1-f_3)>0$, and the inequality persists if we change $\theta_1^L$ to $\theta_1$. As
$f_1-f_3=f_2-f_4+2f_{23}$, it becomes
$(\theta_1-\theta_3-\theta_2+\theta_4)(f_2-f_4)-2(\theta_2-\theta_4)f_{23}>0$, which may be written as
\begin{equation}\label{c2}
\frac{f_{23}}{\pi-(\theta_2+\theta_3)}<\frac{f_2-f_4}{\theta_2-\theta_4}
\end{equation}
because $\theta_2-\theta_4>0$ and $2\pi-2(\theta_2+\theta_3)=\theta_1-\theta_3-\theta_2+\theta_4>0$. On the other hand,
$$\frac{f(\pi)-f_{23}}{\pi-(\theta_2+\theta_3)}<\frac{f_{23}-f_{34}}{(\theta_2+\theta_3)-(\theta_3+\theta_4)}=-\frac{f_{2}-f_{4}}{\theta_2-\theta_4}$$
due to the concavity of $f$, which, as $f(\pi)=0$, contradicts (\ref{c2}). $\Box$

{\bf Proposition 2.} If in a solution of (\ref{cc}) two consecutive
angles are equal, then the two remaining angles are equal. The
configuration has an axis of symmetry.

{\sl Proof.} We assume $\theta_1=\theta_2$ and $\theta_4>\theta_3$,
and deduce a contradiction. By (\ref{cc}), $\theta_1=\theta_2$
implies $-f_{14}=f_{23}=f_4-f_3=0$. But there are only three roots
of $f(\theta)=0$, namely $\pi/3$, $\pi$ and $5\pi/3$. As
$\theta_2+\theta_3=2\pi-(\theta_1+\theta_4)$ and
$\theta_4>\theta_3$, the only possibility is
$\theta_1+\theta_4=5\pi/3$ and $\theta_2+\theta_3=\pi/3$. The last
equality gives us 3 complementary cases, that is, either
$\theta_1=\theta_2=\theta_3=\pi/6$, or
$\theta_3<\pi/6<\theta_1=\theta_2<\pi/3$, or
$\theta_1=\theta_2<\pi/6<\theta_3<\pi/3$. In the first case, we have
$\theta_4=3\pi/2$ and $4(f_4-f_3)=3(2(\sqrt{2}-1)+\sqrt{6})>0$,
therefore the configuration is not central\footnote{Interestingly, if we take the
interaction corresponding to $p=-0.101834199...$ in ({\ref{fp}}),
this configuration is central and is the continuation of ${\cal E}_2$.}. The second case is
impossible because we have both $f_2>f_3$ due to Property 2, and
$f_3-f_2=f_{12}>0$ due to both (\ref{cc}) and
Property 2. In the third case, these two
inequalities are reversed, giving again a contradiction. $\Box$

{\bf Proposition 3.} If in a solution of (\ref{cc}) a diagonal of
the quadrilateral formed by the four small bodies passes through the center of the circumcircle, this diagonal is an axis of symmetry for the configuration.

{\sl Proof.} We assume
$\theta_1+\theta_4=\theta_2+\theta_3=\pi$. Obviously, the four
$\theta_i$'s  belong to $]0,\pi[$, and we have by (\ref{cc})
$f_1-f_2=f_4-f_3=f_{23}=0$, that is, both $f_1=f_2$ and $f_3=f_4$. Therefore, if we have $\theta_1\neq\theta_2$, then
$\theta_3\neq\theta_4$, and by the corollary of Lemma 2, we have both
$\theta_1+\theta_2>2\theta_c$ and $\theta_3+\theta_4>2\theta_c$.  We check that $f'(\pi/2)>0$, which implies by Property 3 that $\pi/2<\theta_c$, which in turn leads to the
contradiction $2\pi=\theta_1+\theta_2+\theta_3+\theta_4>2\pi$. $\Box$

{\sl Remarks.} Proposition 3 corresponds to a case of Proposition 11
in [CLO]. Our method of proof is faster. Our 3 propositions prove
that any solution of (\ref{cc}), i.e. any relative equilibrium of
four separate identical satellites, possesses some symmetry. One
possibility is the equality of two non-adjacent angles, namely
$\theta_1=\theta_3$ or $\theta_2=\theta_4$. In this case, there is
an axis of symmetry that contains only the central body, i.e.\ the
center of the circumcircle. The other possibilities left after
Proposition 1 are the hypotheses of Propositions 2 and 3, and they
both imply the existence of an axis of symmetry passing through the
central body and two of the four small bodies. In the next section
we recall, after [CLO], that there are two configurations in both
classes of symmetry, one of them, the square, being common to both
classes.

\bigskip
\centerline{\bf 4. The symmetric configurations}

{\bf Proposition 4.} [CLO] There are exactly two relative equilibria  (labeled ${\cal E}_1$ and ${\cal E}_3$ on Figure 1) of four separate identical satellites which are symmetric with respect to a diagonal of the quadrilateral of the small bodies.

This is Propostion 12 of [CLO]. The next result is also stated in
[CLO], but we decided to give here a complete proof for several
reasons. First, in several papers from [Alb], the final arguments
are only sketched. Some floating point approximations are used to
discard some solutions, and the authors do not make clear if they
consider that the work left to get a proof is just a routine work,
or if they are satisfied with this degree of rigor. Second, as this
method of proof may be tested on more and more complex situations,
it is important to present what we consider as the most efficient
way to produce a rigorous proof. For example, we choose instead of
Sturm's algorithm the method of rational substitutions, inspired
from Vincent's method, as advised to us by Eduardo Leandro (see
[Lea]). Third, compared to [CLO], we could reduce the degree of the
key polynomial from degree 312 to degree 78.

{\bf Proposition 5.}  [CLO] There are exactly two relative equilibria  (labeled ${\cal E}_2$ and ${\cal E}_3$ on Figure 1) of four separate identical satellites which are symmetric
with respect to an axis which is not a diagonal of the quadrilateral of the small bodies.

{\sl Proof.} We suppose the symmetry $\theta_1=\theta_3$. As we have
the relation $2\theta_1+\theta_2+\theta_4=2\pi$, we need two angles
to get the configuration. If we exchange the values of $\theta_2$
and $\theta_4$, we get essentially the same configuration. So it is
a good idea to choose angles which behave well under this exchange.
We choose $\sigma=(\theta_2+\theta_4)/4$ and
$\nu=(\theta_2-\theta_4)/4$, so that $\theta_1=\pi-2\sigma$,
$\theta_2=2(\sigma+\nu)$, $\theta_4=2(\sigma-\nu)$,
$\theta_1+\theta_2=\pi+2\nu$. A good expression for $f$ is
$$f(\theta)={1\over 4}\cos{\theta\over 2}\sin^{-2}{\theta\over
2}\Bigl(8\sin^3{\theta\over 2}-1\Bigr)$$ This is (\ref{fun2}) rather
than (\ref{fun}), so this expression is not correct if
$\theta\in]2\pi,4\pi[$. We will simply ignore the solutions
corresponding to this interval. Setting $C=\cos\sigma$,
$S=\sin\sigma$, $c=\cos\nu$, $s=\sin\nu$, we get the rational
expressions
$$4f_1=SC^{-2}(8C^3-1),\quad 4f_{12}=-sc^{-2}(8c^3-1)$$
$$4f_2=(Cc-Ss)(Sc+Cs)^{-2}(8(Sc+Cs)^3-1)$$
$$4f_4=(Cc+Ss)(Sc-Cs)^{-2}(8(Sc-Cs)^3-1)$$
The two equations $f_2+f_4-2f_1=0$ and $2f_{12}+f_2-f_4=0$
characterise the central configurations. We denote by $P$ and $sQ$
the respective numerators of the fractions $f_2+f_4-2f_1$ and
$2f_{12}+f_2-f_4$. The reader may use a computer to deduce
expressions of $P$ and $Q$, and continue the computations using
these expressions. We mention here simplified expressions which are
not easy to obtain directly from a computer, due to our
opportunistic use of the relations $C^2+S^2=1$, $c^2+s^2=1$:
$$P=S(C^2-c^2)^2(1-16s^2C^3)-C^3c(s^2+S^2)$$
$$Q=(C^2-c^2)^2(1-16S^2c^3)+Sc^2(C^2+c^2)$$
Our system is $P=sQ=0$, $c^2+s^2=C^2+S^2=1$. This system possesses many complex solutions. We are interested in the solutions such that $S,C,s,c$ are real. Then the $\theta_i$ are real. As we said, we should furthermore impose $0<\theta_i<2\pi$. This condition gives $\sigma>0$ and, together with $\theta_1=\pi-2\sigma$, $\sigma<\pi/2$, i.e.\ $S>0$ and $C>0$.

A first branch of solutions corresponds to $s=0$. We get $c=\pm1$, and
$P=S^2(S^3-cC^3)=0$. The inequalities above allow only the solution $S=C=1/\sqrt2$, $c=1$. This gives $\theta_1=\theta_2=\theta_3=\theta_4=\pi/2$, the square solution ${\cal E}_3$.

Now we consider the branch $s\neq 0$. We
first notice that there are only even powers of $s$ in the
simplified system. The reason is that exchanging $\theta_2$ and
$\theta_4$ changes the sign of $s$, keeping $c$ and $s^2$ invariant.
We replace $s^2$ by $1-c^2$, set $C=(1-t^2)/(1+t^2)$, $S=2t/(1+t^2)$
and consider the numerators
$P_6(t,c)=-32t(t^2-1)^3(1+t^2)^4c^6+\cdots$ and
$Q_7(t,c)=-64t^2(1+t^2)^4c^7+\cdots$ of $P$ and $Q$. Their resultant
in $c$ is a polynomial in $t$ with integer coefficients,
$R=-524288t^5(t^2-1)^{16}(1+t^2)^{32}R_{78}$. The irreducible monic
factor $R_{78}$ is
$$R_{78}=t^{78}+12t^{77}+77t^{76}+464t^{75}-64992t^{74}+\cdots-104961536141944t^{39}+$$
$$+\cdots+77790t^4-464t^3-77t^2-12t-1.$$
As $t=\tan(\sigma/2)$ and  $0<\sigma<\pi/2$, $0<t<1$ and the acceptable roots of $R(t)$ are in the interval $]0,1[$. Substituting successively
$$t={y\over 1+3y},\quad t={1+7y\over 3+20y},\quad t={7(1+y)\over 10(2+y)},\quad t={7+y\over 10+y},$$
in $R_{78}$, we find numerators with respectively 0, 1, 1, 0 variations
of sign, which according to Descartes rule of sign proves that
$R_{78}$ has only two real roots $t_1$ and $t_2$ in $]0,1[$,
satisfying ${1\over 3}<t_1<{7\over 20}<t_2<{7\over 10}$. In order to
prove that $t_1$ does not correspond to a real central
configuration, we need to express the variable $c$ as a function of
$t$. There is a unique expression of $c$ as a polynomial in $t$ of
degree lower than 78 and with rational coefficients, but this
expression is not suitable for practical computations. Many
expressions of $c$ as a rational function of $t$ with integer
coefficients behave well. We can compute the Sylvester matrix of $P_6(t,c)$ and $Q_7(t,c)$ in $c$.
It is a $13\times13$ matrix whose entries are polynomials in $t$
with integer coefficients, and whose determinant is the resultant $R(t)$. We can set for example
$c=F(t)=-M_{1,7}/M_{1,8}$, where $M_{i,j}$ is the minor determinant
of the Sylvester matrix obtained by removing the $i$-th line and the
$j$-th column. We get
$$F(t)={(t+1)(t^2+1)^4N(t)\over 8t^2(t-1)^2D(t)}$$
$$ N(t)=t^{56}+8t^{55}+42t^{54}+248t^{53}-48461t^{52}+\cdots+284953591140
t^{28}+$$
$$+\cdots+47947t^4-248t^3-42t^2-8t-1 $$
$$
D(t)=15t^{59}+7t^{58}+83t^{57}+\cdots+807713099949204
t^{30}+\cdots-141t^2-25t-17
$$
The function $F(t)$ is well-defined and decreasing
from $1/4$ to $1/2$, as may be proved by applying the substitution
{$t=(1+y)/(4+2y)$, respectively to $D(t)$ and to the
numerator of $F'(t)$. So on the interval $]1/3,7/20[$ we have
$c=F(t)>F(7/20)>1$. But $c=\cos(\nu)\leq 1$ so the root $t_1$ should
be rejected. Finally, there is only one solution in this branch satisfying the reality conditions,
which corresponds to the root $t_2$. This solution is ${\cal E}_2$. QED

The reader may use the above technique to prove rigorously many inequalities concerning the solution ${\cal E}_2$, for example, the inequalities in our Theorem. He can check first that $t_2 \in
]3/5,7/10[$. On this interval $F(t)$ is increasing, from $F(3/5)>0$  to $F(7/10)<1$.

The solution ${\cal E}_2$ was described by [Hal] and [SaY], who both gave the estimates $\theta_1=\theta_3=41.5^0$, $\theta_2=37.4^0$, $\theta_4=239.6^0$. It was identified as a linearly stable relative equilibria in [SaY] and [Moe].

{\sl Acknowledgment.} We wish to thank Fathi Namouni, Stefan Renner, Bruno Sicardy and Carles Sim\'o for instructive discussions.

\bigskip

\centerline{\bf References}

[Alb] A. Albouy, {\sl The symmetric central configurations of four equal masses}, Contemporary Mathematics 198 (1996) pp.\ 131--135

[AlL] A. Albouy, J. Llibre, {\sl Spatial central configurations for the 1+4 body problem}, Contemporary Mathematics 292 (2002) pp.\ 1--16

[Alm] A. Almeida Santos, {\sl  Dziobek's configurations in
restricted problems and bifurcation}, Celestial Mechanics and Dynamical
Astronomy 90 (2004) pp.\ 213--238

[AlV] A. Almeida Santos, C. Vidal, {\sl Symmetry of the restricted $4+1$ body problem with equal masses}, Regular and Chaotic Dynamics 12 (2007) pp.\ 27--38

[CLN] J. Casasayas, J. Llibre, A. Nunes, {\sl Central configuration of the planar $1+n$-body problem}, Celestial Mechanics and Dynamical Astronomy 60 (1994) pp.\ 273--288

[CLO] J.M. Cors, J. Llibre, M. Oll\'e, {\sl Central configurations of the planar coorbital satellite problem}, Celestial Mechanics and Dynamical Astronomy 89 (2004) pp.\ 319--342

[CLO2] J.M. Cors, J. Llibre, M. Oll\'e, {\sl On the central configurations of the planar coorbital satellite problem}, lecture by J. Llibre in the third Tianjin international conference on nonlinear analysis. Hamiltonian systems and celestial mechanics. June 16-20, 2004

[Dan] J.M.A. Danby, {\sl The stability of the triangular Lagrangian point in the general problem of three bodies}, Astronomical Journal 69 (1964) pp.\ 294--296

[Gas] G. Gascheau, {\sl Examen d'une classe d'\'equations diff\'erentielles et application \`a un cas particulier du probl\`eme des trois corps}, Comptes rendus hebdomadaires 16 (1843) pp.\ 393--394

[Hal] G.R. Hall, {\sl Central configurations in the planar $1+n$-body problem}, Preprint, Boston University

[HuS] X. Hu, S. Sun, {\sl Morse Index and Stability of Elliptic Lagrangian Solutions in the Planar Three-Body Problem}, preprint (2008)

[Lag] J.L. Lagrange, {\sl Essai sur le probl\`eme des trois corps}, \oe uvres v.6 (1772) pp.\ 229--324

[Lea] E.S.G. Leandro, {\sl Bifurcation and stability of some symmetrical classes of central configurations}, Thesis, University of Minnesota (2001)

[Max] J.C. Maxwell, {\sl Stability of the motion of Saturn's rings}, (1859) in {\sl The Scientific Papers of James Clerk Maxwell}, Cambridge University Press (1890) or {\sl Maxwell on Saturn's Rings}, MIT Press (1983)

[MeS] K. Meyer, D. Schmidt, {\sl Elliptic relative equilibria in the N-body problem}, J.\ Differential Equations 214 (2005) pp.\ 256--298

[Moe] R. Moeckel, {\sl Linear stability of relative equilibria with a dominant mass}, Journal of Dynamics and Differential Equations 6 (1994) pp.\ 37--51

[Moe2] R. Moeckel, {\sl Relative equilibria with clusters of small masses}, Journal of Dynamics and Differential Equations 9 (1997) pp.\ 507--533

[MS] R. Mart\'\i nez, C. Sim\'o, {\sl Homoclinic and heteroclinic connections related to the center manifold of collinear libration points in the 3D RTBP}, lecture by C. Sim\'o in Journ\'ees en l'honneur d'Alain Chenciner, Paris (2003)

[MSS] R. Mart\'\i nez, A. Sam\`a, C. Sim\'o, {\sl Stability of homographic solutions of the planar three-body problem with homogeneous potentials}, International conference on Differential equations, Hasselt 2003, Dumortier, Broer, Mawhin, Vanderbauwhede and Verduyn Lunel eds, World Scientific (2004) pp.\ 1005--1010

[NaP] F. Namouni, C. Porco, {\sl The confinement of Neptune's ring arcs by the moon Galatea}, Nature 417 (2002) pp.\ 45--47

[ONe] K. O'Neil,  {\sl Stationary configurations of point vortices}, Trans.\  Amer.\  Math.\  Soc.\
302 (1987) pp.\ 383--425

[PoS] G. P\'olya, G. Szeg\"o, {\sl Problems and Theorems in Analysis}, translation C.E.\ Billigheimer, Springer-Verlag (1976)

[Ren] S. Renner, {\sl Dynamique des anneaux et des satellites plan\'etaires~: application aux arcs de Neptune et au syst\`eme Prom\'eth\'ee-Pandore}, Th\`ese, Observatoire de Paris (2004) tel.archives-ouvertes.fr/tel-00008103/

[ReS] S. Renner, B. Sicardy, {\sl Stationary configurations for co-orbital satellites with small arbitrary masses}, Celestial Mechanics and Dynamical Astronomy 88 (2004) pp.\ 397--414

[Rob] G. Roberts, {\sl Linear stability of the  $1+n$-gon relative equilibrium}, Hamiltonian systems and celestial mechanics (HAMSYS-98), World Scientific Monographs Series in Maths 6 (2000) pp.\ 303--330

[Rob2] G. Roberts, {\sl Linear stability of the elliptic Lagrangian triangle solutions in the three-body problem}, J.\ Differential Equations 182 (2002) pp.\ 191--218

[RoS] P. Robutel, J. Souchay, {\sl An introduction to the dynamics of the Trojan asteroids}, in {\sl Recent Investigations in the Dynamics of  Celestial Bodies in the Solar and Extrasolar Systems},  Souchay and Dvorak eds, Lecture notes in physics, Springer (2009) 

[SaH] H. Salo, J. H\"anninen, {\sl Neptune's partial rings: action of Galatea on self-graviting arc particles}, Science 282 (1998) pp.\ 1102--1104

[SaY] H. Salo, C.F. Yoder, {\sl The dynamics of coorbital satellite systems}, Astronomy and Astrophysics 205 (1988) pp.\ 309--327

[SiR] B. Sicardy, S. Renner, {\sl On the confinement of Neptune's ring arcs by co-orbital moonlets (abstract)}, Bulletin of the American Astronomical Society 35 (2003) p.\ 929

[Xia]  Z. Xia, {\sl Central configurations with many small masses}, J.\ Differential Equations 91 (1991)
pp.\ 168--179

\end{document}